\documentclass[article,amsmath,amssymb]{revtex4-2}
\usepackage{graphicx}
\usepackage{dcolumn}
\usepackage[colorlinks]{hyperref}
\usepackage{bm}

\begin{document}

\title{Cooperative decay of an ensemble of atoms in a one-dimensional chain with a single excitation}

\author{Nicola Piovella}
\affiliation{Dipartimento di Fisica "Aldo Pontremoli", Universit\`{a} degli Studi di Milano, Via Celoria 16, I-20133 Milano, Italy \&
INFN Sezione di Milano, Via Celoria 16, I-20133 Milano, Italy}

\begin{abstract}
We propose a new expression of the cooperative decay rate of a one-dimensional chain of $N$ two-level atoms in the single-excitation configuration.  From it, the interference nature of superradiance and subradiance arises naturally, without the need of solving the eigenvalue problem of the atom-atom interaction Green function. The cooperative decay rate can be interpreted as the imaginary part of the expectation value of the effective non-Hermitian Hamiltonian of the system, evaluated over a generalized Dicke state of $N$ atoms in the single-excitation manifold. Whereas the subradiant decay rate is zero for an infinite chain, it decreases as $1/N$ for a finite chain.  A simple approximated expression for the cooperative decay rate is obtained as a function of the lattice constant $d$ and the atomic number $N$.  The results are obtained first for the scalar model and then extended to the vectorial light model, assuming all the dipoles aligned.
\end{abstract}

\maketitle

\section{Introduction}

Cooperative spontaneous emission by $N$ excited two-level atoms has been extensively studied, since the seminal work by Dicke in 1954 \cite{Dicke1954} and Lehmberg in 1970 \cite{Lehmberg1970}. Whereas superradiance, i.e. enhanced spontaneous emission  due to positive interference between the emitters, has been well understood \cite{Bonifacio1971,Gross1982}, subradiance, i.e. inhibited emission due to negative interference between the emitters, is more elusive and difficult to observe \cite{Crubellier1985,Bienaime2012,Guerin2016}. Nevertheless, the number of studies on subradiance  has seen a large increase in the last few years, as it offers the opportunity of storing photons in emitter ensembles for times longer than the single emitter lifetime \cite{Scully2015,Jen2016,Facchinetti2016,Bettles2016,Asenjo2017,Needham2019,Cech2023}.

In disordered systems, the cooperative decay must be studied numerically, usually by solving the dynamics of an initially excited ensemble \cite{Bienaime2013}. Eventually, in the linear regime of weak excitation, useful informations can be obtained from the spectrum of the eigenvalues of the system \cite{Bellando2014}, determined numerically by diagonalizing the finite matrix associated to the Green operator describing the coupling between the emitters. Things are apparently more simple when atoms form ordered arrays, easier to treat theoretically. This is also the reason why propagation of excitations in lattices with different dimensionalities has been generally investigated more than in disordered systems. In fact, the existing literature on interaction of coherent light with atomic lattices is very reach \cite{Zoubi2010,Jenkins2012,Bettles2015,Masson2020,Masson2022}. Most of it presents a 'solid-state physics' point of view, investigating the transport of particular photonic (but not necessarily) modes through ordered samples, often taken infinite but in few exception taken also finite. For instance, infinite and finite chains of two-level atoms have been considered in ref. \cite{Asenjo2017}. Conversely, superradiance and subradiance are typically interpreted as quantum optics phenomena and are seen associated to cooperativity in the emission or light scattering of  ensemble of atoms \cite{Gross1982,Scully2015}. The bridge between these two complementary descriptions is not yet completely established.

The aim of this paper is to add some new insight on the study of a one-dimensional array of atoms in the single-excitation configuration, i.e. with only one atom excited among $N$. Starting from the effective non-Hermitian Hamiltonian, which includes an imaginary part describing the cooperative spontaneous decay and a real part describing the cooperative energy shift \cite{Lehmberg1970,Friedberg1973}, we focus on the cooperative decay only, without solving the eigenvalue spectrum problem, but adopting a different approach: we introduce a collective decay function $\Gamma_k$, parametrized by a continuous label $k$, which in the case of infinite chain represents the Fourier transform of the system. For a finite chain, the collective decay $\Gamma_k$ is able to catch the main characteristics of the superradiant and subradiation emission by the excited atoms in the chain, enlightening  the symmetry properties of the $N$-atoms state associated to such cooperative phenomena. In particular, starting from the solution for an infinite chain, obtained as a limit case of the general solution, we obtain an approximated analytical expression for the cooperative decay from a finite chain of $N$ atoms when $N$ is large. The analysis is carried on initially assuming the scalar model of light and neglecting the vectorial nature of the dipoles. The scalar model is particular attractive because, since the polarization direction does not play any role, is able to catch to main features of cooperativity just considering the relative phases of the emitters, taken at a fixed distance along the chain. So, the control parameters are in this case only the lattice constant $d$ and the atom number $N$. Then, we extend the results to the vectorial light model for a set of $N$ equally oriented dipoles. We see that the scalar light model still provides a good approximation of the vectorial light model, either when the directions of the dipoles are randomized or in the limit of large lattice constant.

The paper is organized as follow. In sec.II we start from the non-Hermitian Hamiltonian in the scalar light model and we introduce the collective decay function $\Gamma_k$. We show how $\Gamma_k$ can be calculated exactly and we recover the solution for the infinite chain. Then, we discuss the collective decay function $\Gamma_k$ for a finite chain and show that it corresponds to the imaginary part of the expectation value of the non-Hermitian Hamiltonian, evaluated over an entangled state, which reduces to the Dicke superradiant and subradiant states for particular values of $k$. In sec. III we extend the results of sec.II to the vectorial light model. Sec.IV is dedicated to the conclusions.

\section{Scalar Model}\label{s:model}

We consider $N$ two-level atoms with the same atomic transition frequency $\omega_0=ck_0$, linewidth $\Gamma$ and dipole $d$. The atoms are prepared in a single-excitation state;  $|g_j\rangle$ and $|e_j\rangle$ are the ground and excited states, respectively, of the $j$-th atom, $j=1,\ldots,N$, which is placed at position $\mathbf{r}_j$.
We consider here the single-excitation effective Hamiltonian in the scalar approximation, whereas the exact vectorial model will be considered in the next section. If we assume that only one photon is present, when tracing over the radiation degrees of freedom the dynamics of the atomic system can be described by the non-Hermitian Hamiltonian \cite{Akkermans2008,Bienaime2013}
\begin{eqnarray}
\hat H & =&-i\frac{\hbar}{2}\sum_{j,m}G_{jm}\,
    \hat{\sigma}_j^\dagger\hat\sigma_m,\label{Heff}
\end{eqnarray}
 where $\hat\sigma_j=|g_j\rangle\langle e_j|$ and $\hat\sigma_j^\dagger=|e_j\rangle\langle g_j|$ are the lowering and raising operators,
\begin{equation}\label{gammajm}
    G_{jm}=
   	\left\{
    \begin{array}{ll}
    \Gamma_{jm}-i\, \Omega_{jm} & \mbox{if}~j\neq m, \\[1ex]
    \Gamma & \mbox{if}~j = m,
\end{array}
	\right.
\end{equation}
and
\begin{equation}\label{gammajm:bis}
    \Gamma_{jm} = \Gamma\frac{\sin(k_0r_{jm})}{k_0r_{jm}}\quad , \quad
    \Omega_{jm} = \Gamma\frac{\cos(k_0r_{jm})}{k_0r_{jm}},
\end{equation}
where $r_{jm}=|\mathbf{r}_j-\mathbf{r}_m|$.
$\hat H$ contains both real and imaginary parts, which takes into account that the excitation is not conserved since it can leave the system by emission. 
We focus our attention on the decay term $\Gamma_{jm}$. It can be obtained as the angular average of the radiation field propagating between the  two atomic positions $\mathbf{r}_j$ and $\mathbf{r}_m$ with wave-vector $\mathbf{k}=k_0(\sin\theta\cos\phi,\sin\theta\sin\phi,\cos\theta)$,
\begin{equation}
\Gamma_{jm}=\frac{\Gamma}{2}\left\langle e^{-i\mathbf{k}\cdot (\mathbf{r}_j-\mathbf{r}_m)} 
+\mathrm{c.c.}\right\rangle_\Omega\label{Sk}
\end{equation}
where the angular average is defined as
\[
\left\langle f(\theta,\phi)\right\rangle_\Omega=\frac{1}{4\pi}\int_0^{2\pi}d\phi\int_0^\pi \sin\theta f(\theta,\phi)d\theta.
\]
Eq.(\ref{Sk}) provides a simple interpretation of $\Gamma_{jm}$ as the coupling between the $j$th atom and the $m$th atom, mediated by the photon shared between the two atoms and averaged over all the vacuum modes. The interaction includes also the energy shift $\Omega_{jm}$, which however we will not consider here, since it is not relevant for our study.
The important technical point is that Eq.(\ref{Sk}) allows to factorize $\Gamma_{jm}$ in the product of two terms, before averaging them over the total solid angle.

Things become particularly simple if we consider $N$ atoms placed along a linear chain with lattice constant  $d$ i.e. with positions $\mathbf{r}_j=d(j-1)\mathbf{\hat e}_z$, with $j=1,\dots ,N$.
Then, we can write
\begin{equation}
e^{-i\mathbf{k}\cdot \mathbf{r}_j}=E_j=e^{-ik_0d(j-1)\cos\theta}
\end{equation}
and
\begin{equation}
\Gamma_{jm}=\frac{\Gamma}{2}\left\langle E_j E_m^*
+\mathrm{c.c.}\right\rangle_\Omega.\label{Gamma:1D}
\end{equation}

\subsection{Collective decay rate}

We introduce the following collective variable,
\begin{equation}
\Gamma_k=\frac{1}{N}\sum_{j=1}^N\sum_{m=1}^N\Gamma_{jm}e^{ikd(j-m)}\label{FT}
\end{equation}
depending on the continuous index $k\in[0,2\pi/d)$. While (\ref{FT}) represents the Fourier transform of $\Gamma_{jm}$ only in the case of the infinite chain, such that the system is periodic, nevertheless the continuous variable $k$ is still a good label for the modes when $N$ is sufficiently large \cite{Asenjo2017}. We refer to $\Gamma_k$ as the 'continuous spectrum' of the decay rate, outlining that we are not referring to the discrete spectrum of the eigenvalues of the system, described by the non-Hermitian Hamiltonian (\ref{Heff}), but we introduce a quantity related to the phase of the collective state of the $N$ atoms emitting a single photon. A more precise interpretation of it will be provided at the end of the section. For the moment, let just proceed in its evaluation. By using Eq.(\ref{Gamma:1D}) in Eq.(\ref{FT}) we can write
\begin{eqnarray}
  \Gamma_k &=&\frac{\Gamma}{N}
  \langle|F_k(\theta)|^2\rangle_\Omega
\end{eqnarray}
where
\begin{eqnarray}
|F_k(\theta)|^2&=&\left|\sum_{j=1}^N e^{i(k-k_0\cos\theta)d(j-1)}\right|^2=\frac{\sin^2[(k-k_0\cos\theta)dN/2]}{\sin^2[(k-k_0\cos\theta)d/2]}
\label{Fk}
\end{eqnarray}
and
\begin{eqnarray}
  \Gamma_k &=&\frac{\Gamma}{4\pi N}\int_0^{2\pi}d\phi\int_0^\pi\sin\theta|F_k(\theta)|^2d\theta
  =\frac{\Gamma}{k_0dN}\int_{(k-k_0)d/2}^{(k+k_0)d/2}\frac{\sin^2(Nt)}{\sin^2 t}dt\label{Gk:1}
\end{eqnarray}
where we changed the integration variable from $\theta$ to $t=(k-k_0\cos\theta)d/2$.
For large $N$, we can approximate, in the integral of Eq.(\ref{Gk:1}),
\begin{equation}
\frac{\sin^2(Nt)}{\sin^2t}\approx N^2\sum_{m=-\infty}^{+\infty} \mathrm{sinc}^2\left[\left(t-m\pi\right)N\right],\label{approx}
\end{equation}
where $\mathrm{sinc}(x)=\sin x/x$,
so that
\begin{eqnarray}
  \Gamma_k &=&
  =\frac{\Gamma N}{k_0d}\sum_{m=-\infty}^{+\infty} \int_{(k-k_0)d/2}^{(k+k_0)d/2}\mathrm{sinc}^2\left[\left(t-m\pi\right)N\right]dt.\label{Gamma:scalar}
\end{eqnarray}

\subsection{Infinite chain}

In the limit $N\rightarrow\infty$,
\begin{equation}
\mathrm{sinc}^2\left[\left(t-m\pi\right)N\right]\rightarrow \frac{\pi}{N}\delta(t-m\pi)\label{Dirac}
\end{equation}
where $\delta(x)$ is the Dirac delta function. Thus,
\begin{eqnarray}
  \Gamma_k &=&\frac{\Gamma\pi}{k_0d}\sum_{m=-\infty}^{+\infty} \Pi[2m\pi-k_0d<kd<2m\pi+k_0d]\label{approx:scalar}
\end{eqnarray}
where $\Pi(a<x<b)$ is the rectangular function, equal to 1 for $a<x<b$ and 0 elsewhere. Eq.(\ref{approx:scalar}) is the solution for an infinite chain, where $k$ is the true index of the modes. The fact that $\Gamma_k$ is a Fourier transform in the continuous variable $k$ is the consequence of the Bloch's theorem. Although the result (\ref{approx:scalar}) is not new (see for instance ref.\cite{Asenjo2017}, obtained analytically using the vectorial model but only for $k_0d<\pi$), it will be useful for our study for two reasons: 
\begin{description}
\item[(a)]  We aim to compare the scalar model with the vectorial model and see what are the most relevant differences; the scalar model is more appealing, since it neglects the vectorial nature of the dipoles composing the chain, focusing only on their phases. Also, the scalar model is not completely unrealistic, since we will see that can be recovered  from the vectorial model assuming the dipoles randomly oriented.
\item[(b)] We are interested to describe a finite chain, with a finite number $N$ of atoms, where the infinite chain represents a limit case of it. Many previous studied on infinite and finite chains (usually based on the numerical evaluation of the complex eigenvalues of the matrix $G_{jm}$ in Eq.(\ref{Heff})) have a 'solid-state physics' approach \cite{Asenjo2017}, investigating the guided propagation of photonic modes along the chain. On the contrary, we are interest to the point of view of the atoms, where superradiance and subradiance arise from positive and negative interference of $N$ emitters, respectively. In the paper by Bettles et al. \cite{Bettles2016} these two points of view are equally well discussed.
\end{description}

\begin{figure}
       \centerline{\scalebox{0.5}{\includegraphics{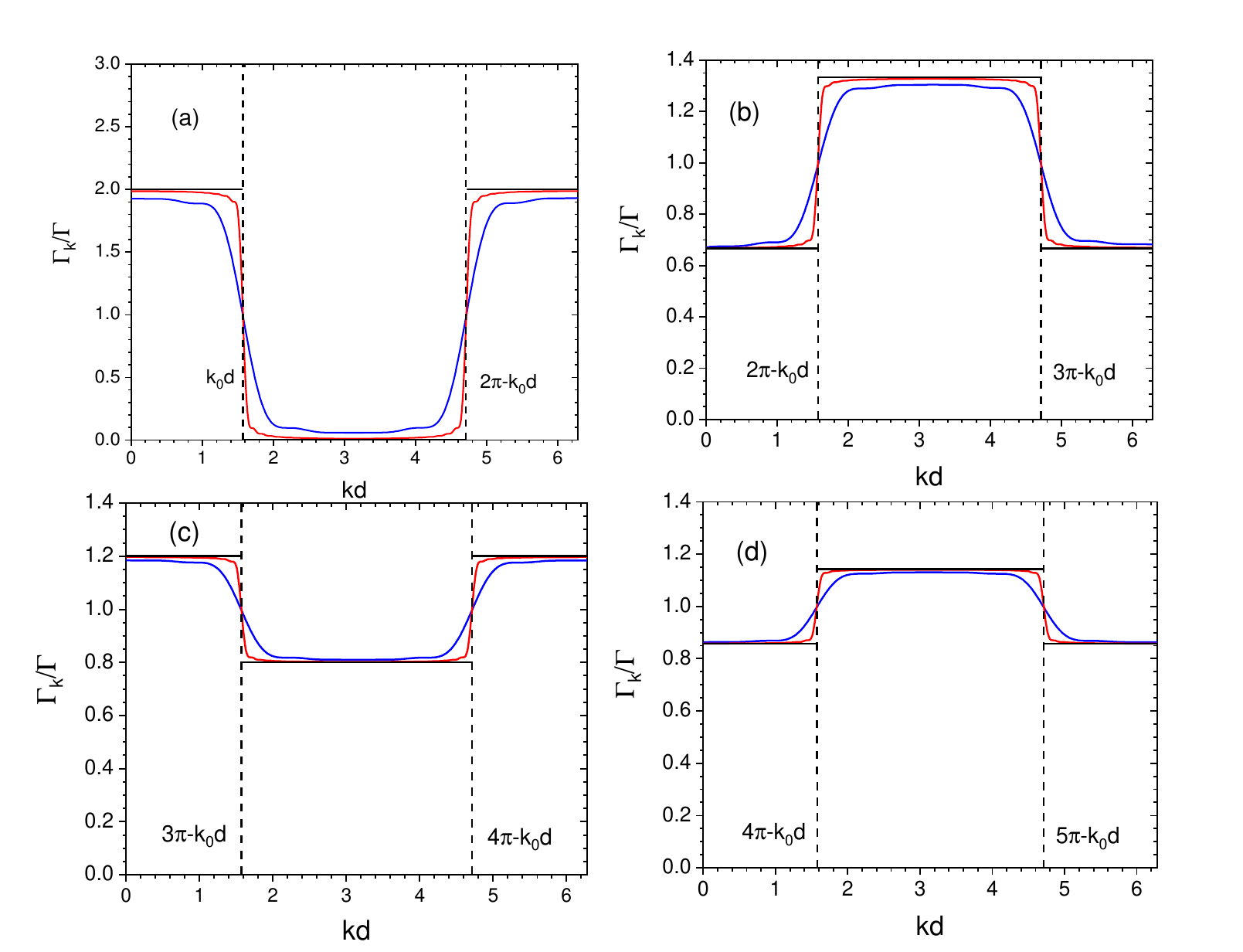}}}
      \caption{ $\Gamma_k/\Gamma$ vs $kd$ for (a) $k_0d=\pi/2$, (b) $k_0d=3\pi/2$, (c) $k_0d=5\pi/2$ and (d) $k_0d=7\pi/2$, obtained from Eq.(\ref{Gamma:scalar}) with $N=10$ (blue line) and $N=50$ (red line). Black line: solution (\ref{approx:scalar}).}
       \label{fig1}
    \end{figure}

From the result (\ref{approx:scalar}) we see that, when $k_0d<\pi$ (i.e. $d/\lambda_0<0.5$), $\Gamma_k=0$ for $k_0<k<2\pi/d-k_0$ i.e. \textit{full subradiance} (see Fig.\ref{fig1}a), whereas for $0<k<k_0$ and $2\pi/d-k_0<k<2\pi/d$, $\Gamma_k=\Gamma(\pi/k_0d)>\Gamma$ i.e. \textit{enhanced radiance} \cite{Asenjo2017,Cech2023}. For larger lattice constant, \textit{slight subradiance} and enhanced radiance alternate every half-wavelength  i.e. at increasing interval $\pi$ of $k_0d$ (see Fig.\ref{fig2} b-d): it results than, when $m\pi<k_0d<(m+1)\pi$, in the interval $(m+1)\pi<(k+k_0)d<(m+2)\pi$, $\Gamma_k=\Gamma(m+1)\pi/k_0 d>\Gamma$ for $m$ odd and $\Gamma_k=\Gamma m\pi/k_0 <\Gamma$ for $m$ even.
In the external intervals $(k+k_0)d<\pi$ and $(k+k_0)d>(m+2)\pi$, $\Gamma_k=\Gamma m\pi/k_0 d <\Gamma$ for $m$ odd and $\Gamma_k=\Gamma(m+1)\pi/k_0 d >\Gamma$ for $m$ even.
Hence, for each subsequent  $\pi$ interval of $k_0d$, two regions of the spectrum can be identified, one where $\Gamma_k$ is less than $\Gamma$ and the other one where $\Gamma_k$ is larger than $\Gamma$. The difference between these two values is $\Delta\Gamma=\pi/k_0d$. The only case of full subradiance, with zero decay rate, occurs for $k_0d<\pi$. 

\subsection{Finite chain}

\begin{figure}
       \centerline{\scalebox{0.5}{\includegraphics{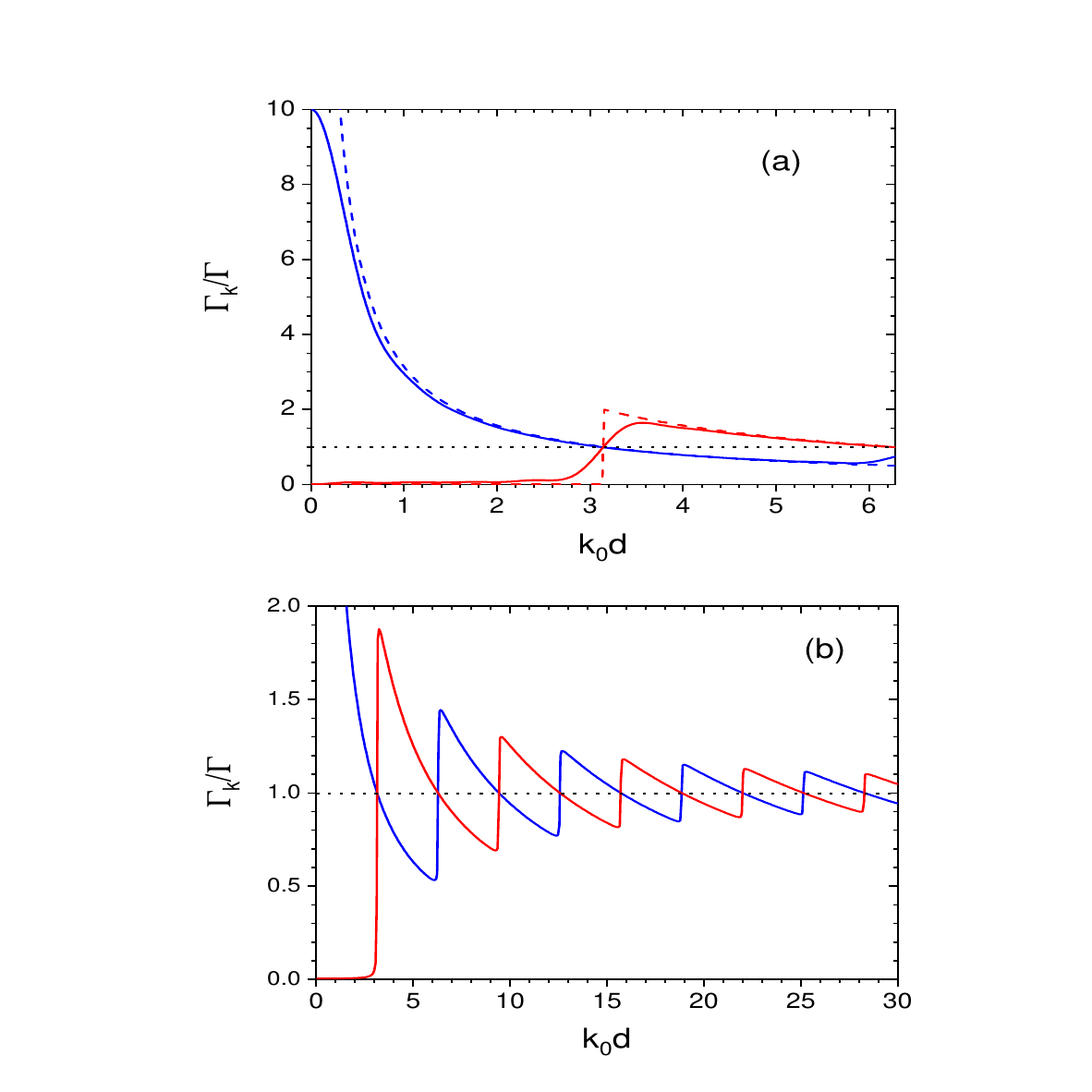}}}
      \caption{ $\Gamma_k/\Gamma$ vs $k_0d$ for $k=0$ (blue line) and $k=\pi/d$ (red line), obtained from Eq.(\ref{Gamma:scalar}) with (a) $N=10$ and (b) $N=100$. The dashed line in (a) is the analytic solution (\ref{approx:scalar}) for $N\rightarrow\infty$.}
       \label{fig2}
    \end{figure}

\begin{figure}
       \centerline{\scalebox{0.3}{\includegraphics{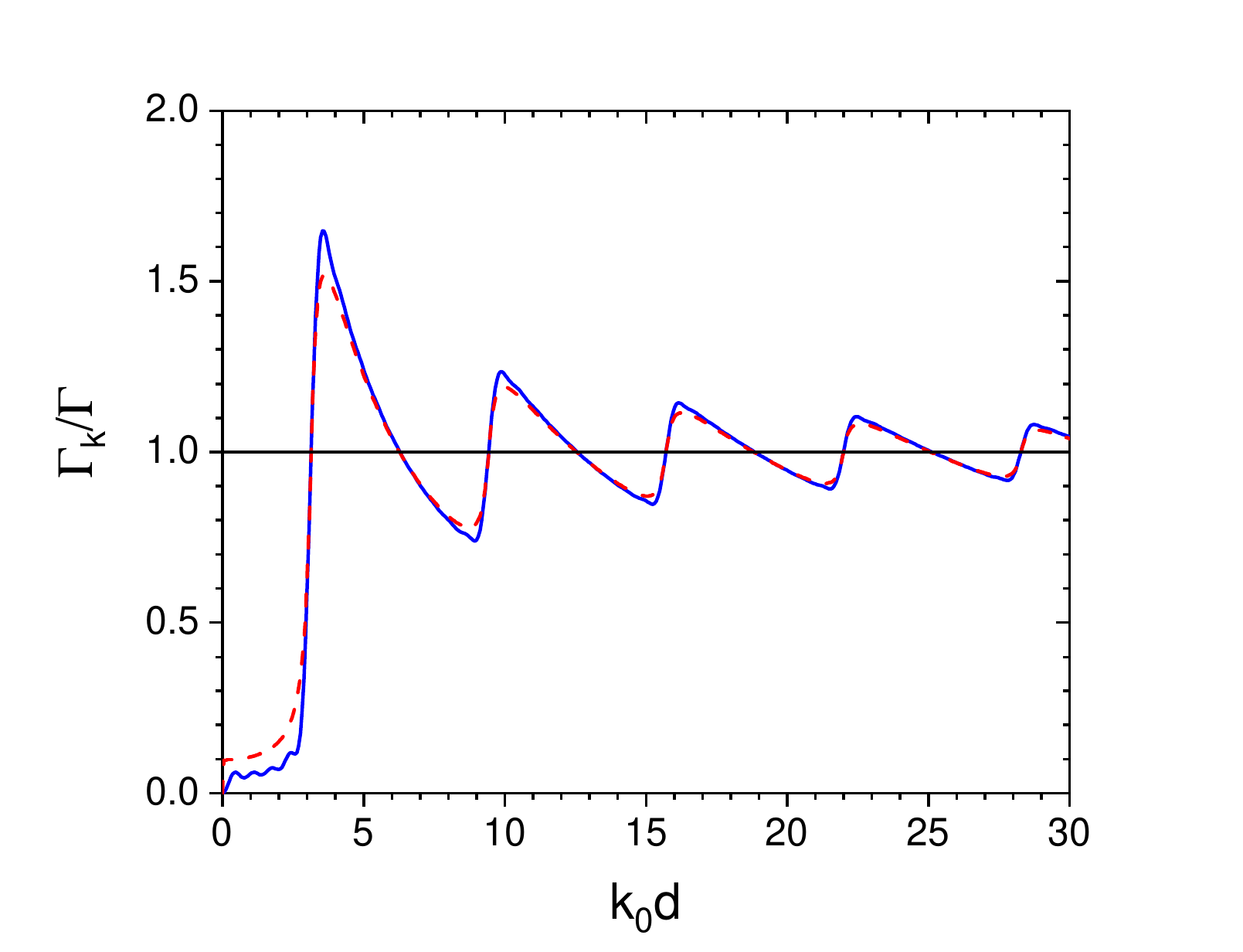}}}
      \caption{ $\Gamma_k/\Gamma$ vs $k_0d$ for $k=\pi/d$ for $N=10$, obtained from exact expression of Eq.(\ref{Gamma:scalar}) (blue continuous line) and from the approximated solution of Eq.(\ref{approx:atan}) (red dashed line).}
       \label{fig3}
    \end{figure}

The case of a finite chain requires the evaluation of the integral in Eq.(\ref{Gamma:scalar}).
Fig.\ref{fig2} shows $\Gamma_k/\Gamma$ vs $k_0d$ for $k=0$ (blue line) and $kd=\pi$ (red line), for $N=10$ (fig.\ref{fig2}a) and $N=100$ (Fig.\ref{fig2}b). The dashed line in Fig.\ref{fig2}a is the analytic solution (\ref{approx:scalar}) for infinite chain. We see that for an infinite chain and $k_0d<2\pi$, $\Gamma_{k=0}=\Gamma(\pi/k_0d)$, whereas for $\pi<k_0d<2\pi$, $\Gamma_{k=\pi/d}=\Gamma(2\pi/k_0d)$.
We have now the elements to discuss in details superradiance and subradiance in a finite chain. Let consider the blue line of Fig.\ref{fig2}a and Fig.\ref{fig2}b, corresponding to $k=0$. For $k_0d\ll 1/N$, $\Gamma_{k=0}\sim \Gamma N$, which corresponds to the Dicke superradiance  of $N$ atoms confined in a region smaller that the optical wavelength, such that they emit all in phase. Increasing $k_0d$, the distance between the atoms increases and they become less correlated: the cooperative emission rate decreases until, at $k_0d=\pi$, it reaches the uncorrelated value $\Gamma$ (see Fig.\ref{fig2}a). Then, beyond $k_0d=\pi$ the interference between the emitters becomes destructive  and $\Gamma_{k=0}$ drops below $\Gamma$ (\textit{'slight' subradiance}), until $k_0d$ reaches the value $2\pi$. At this value of the lattice constant $\Gamma_{k=0}$ has a rapid jump from the minimum $\Gamma(\pi/k_0d)$ to the maximum $\Gamma(3\pi/k_0d)$ (see Fig.\ref{fig2}b, blue line). Then, the process repeats itself, with a jump of $\Gamma(m+1)/m$ each time $k_0d$ crosses the value $m(2\pi)$.
Conversely, subradiance is described by the red line of Fig.\ref{fig2}a,b, with $kd=\pi$: when $k_0d\ll 1/N$, nearest neighbor atoms emit with opposite phases and spontaneous emission is inhibited. Surprisingly, this negative interference is preserved also for larger values of the lattice constant, up to $k_0d=\pi$. Then, beyond $k_0d=\pi$, we observe an alternation of enhanced and inhibited emission, mirroring the behavior of $\Gamma_{k=0}$. Although full subradiance, $\Gamma_k=0$, has a transparent interpretation as modes guided without radiative losses beyond the light line $k=k_0$ \cite{Asenjo2017}, however the interpretation 'from the point of view of the atoms', in terms of negative interference between the emitters in a finite chain, is less intuitive, as discussed for instance in ref.\cite{Bettles2016}. To this aim, we need to consider the \textit{superradiant} and \textit{subradiant} states of the system as originally defined by Dicke \cite{Dicke1954}. Before treating them in the next subsection, it is useful to obtain further analytic expressions of the collective decay rate $\Gamma_k$ for the finite chain. An approximated analytic expression can be obtained replacing the function $\mathrm{sinc}^2(x)$ in Eq.(\ref{Gamma:scalar}) with the Lorentzian function $1/(1+x^2)$, which has the same peak value of unity and the same normalization  value of $\pi$. Then, the integral in Eq.(\ref{Gamma:scalar}) yields
\begin{eqnarray}
  \Gamma_k &=&\frac{\Gamma}{k_0d}\sum_{m=-\infty}^{+\infty} \left[
  \arctan(b_m)-\arctan(a_n)\right]\label{approx:atan}
\end{eqnarray}
where $a_m=[(k-k_0)d/2-m\pi]N$ and $b_m=[(k+k_0)d/2-m\pi]N$. Eq.(\ref{approx:atan}) approximates well the behavior obtained from the exact result of Eq.(\ref{Gamma:scalar}), as shown in Fig.\ref{fig3} for $k=\pi/d$ and $N=10$.

 This expression allows to evaluate the subradiant decay rate in the limit of large $N$, using the identity $\arctan(z)=\pm \pi/2-\arctan (1/z)$, where the positive sign is for $z>0$ and the negative sign for $z<0$. Then, for
$k_0d<\pi$ and the band interval $k_0<k<2\pi/d-k_0$, the main contribution in the sum in Eq.(\ref{approx:atan}) comes for the terms $m=0$ and $m=1$ which, for large $N$ and far from the interval edges, yields
\begin{eqnarray}
  \Gamma_k &\approx &\frac{4\Gamma}{N}\left[\frac{1}{(kd)^2-(k_0d)^2}+\frac{1}{(kd-2\pi)^2-(k_0d)^2}\right].\label{approx:sub}
\end{eqnarray}
We observe the dependence on $1/N$, typical of subradiance \cite{Bienaime2012}.
In particular, at the center of the band, $k=\pi/d$, and for $k_0d=\pi/2$, $\Gamma_{k}\approx(32/3\pi^2N)\Gamma$. 
Conversely, superradiance is obtained from Eq.(\ref{approx:atan}) for $k=0$ and $m=0$,
\begin{equation}
\Gamma_{k=0}=\frac{2\Gamma}{k_0d}\arctan\left(\frac{k_0dN}{2}\right).
\end{equation}
In the limit $k_0d\ll 1/N$, $\Gamma_{k=0}\approx \Gamma N[1-(k_0d N)^2/12+\dots]$, as it can be observed in Fig.\ref{fig2}a.

\subsection{Symmetric and anti-symmetric states}

It is easy to prove that
\begin{equation}
\Gamma_k=-\frac{2}{\hbar}\mathrm{Im}\langle k|\hat H|k\rangle,
\end{equation}
where $\hat H$ is the non-Hermitian Hamiltonian (\ref{Heff}) and 
\begin{equation}
|k\rangle=\frac{1}{\sqrt{N}}\sum_{j=1}^N e^{ikd(j-1)}|j\rangle,\label{Dk}
\end{equation}
where $k\in [0,2\pi/ d)$ and $|j\rangle=|g_1,\dots,e_j,\dots,g_N\rangle$.
For $k=0$, the state (\ref{Dk}) corresponds to the Dicke state \cite{Dicke1954}, whereas for $k=k_0$ is the Timed-Dicke state introduced by Scully and coworkers \cite{Scully2006}, corresponding to the entangled state of $N$ atoms, where a single photon with momentum $\hbar k_0$ along the axis chain has been absorbed. For $kd=\pi$ the state (\ref{Dk}) is subradiant, since nearest-neighbor atoms have opposite phases. This picture is in agreement with the analysis of ref.\cite{Bettles2016}, where the nearest-neighbor phase difference  $\phi^\ell_{i+1}-\phi^\ell_i$ for $i=1,\dots,N$ is determined for the $N$ eigenmodes of the system, with index $\ell=1,\dots,N$ ordered from the largest to the smaller decay rate. In our case, the nearest-neighbor phase difference for the state $|k\rangle$ is $kd$, which provides a simple interpretation of enhancement or inhibition of spontaneous emission by $N$ in-phase or out-of-phase atoms in the chain.  
It is interesting to observe that for an infinite chain and in the scalar model the imaginary part of the expectation value 
$\langle k|\hat H|k\rangle$ is constant. This is not true for a finite chain, but it is however a very good approximation for $N$ sufficiently large (see Fig.\ref{fig1}a). Moreover, the fact that $\Gamma_k$ is independent on $k$ for the infinite chain is a consequence of the scalar model, for which the spontaneous emission in the vacuum modes is isotropic. This is not true for the vectorial model, as we will see in the next section. 

It is interesting to note that the states $|k\rangle$ form a complete basis of the single-excitation manifold. In fact
\begin{equation}
\frac{d}{2\pi}\int_0^{2\pi/d}dk |k\rangle\langle k|=\sum_{j=1}^N |j\rangle\langle j|=1.
\end{equation}
As expected, the states $|k\rangle$ are not orthogonal for a finite chain, since
\begin{equation}
\langle k'|k\rangle=\frac{\sin[(k-k')dN/2]}{\sin[(k-k')d/2]}e^{i(k-k')d(N-1)/2},
\end{equation}
but they become so for an infinite chain, $\langle k'|k\rangle\rightarrow \delta(k-k')$ for $N\rightarrow\infty$.

\subsection{Spectrum and eigenvalues}   
\begin{figure}
      \centerline{\scalebox{0.5}{\includegraphics{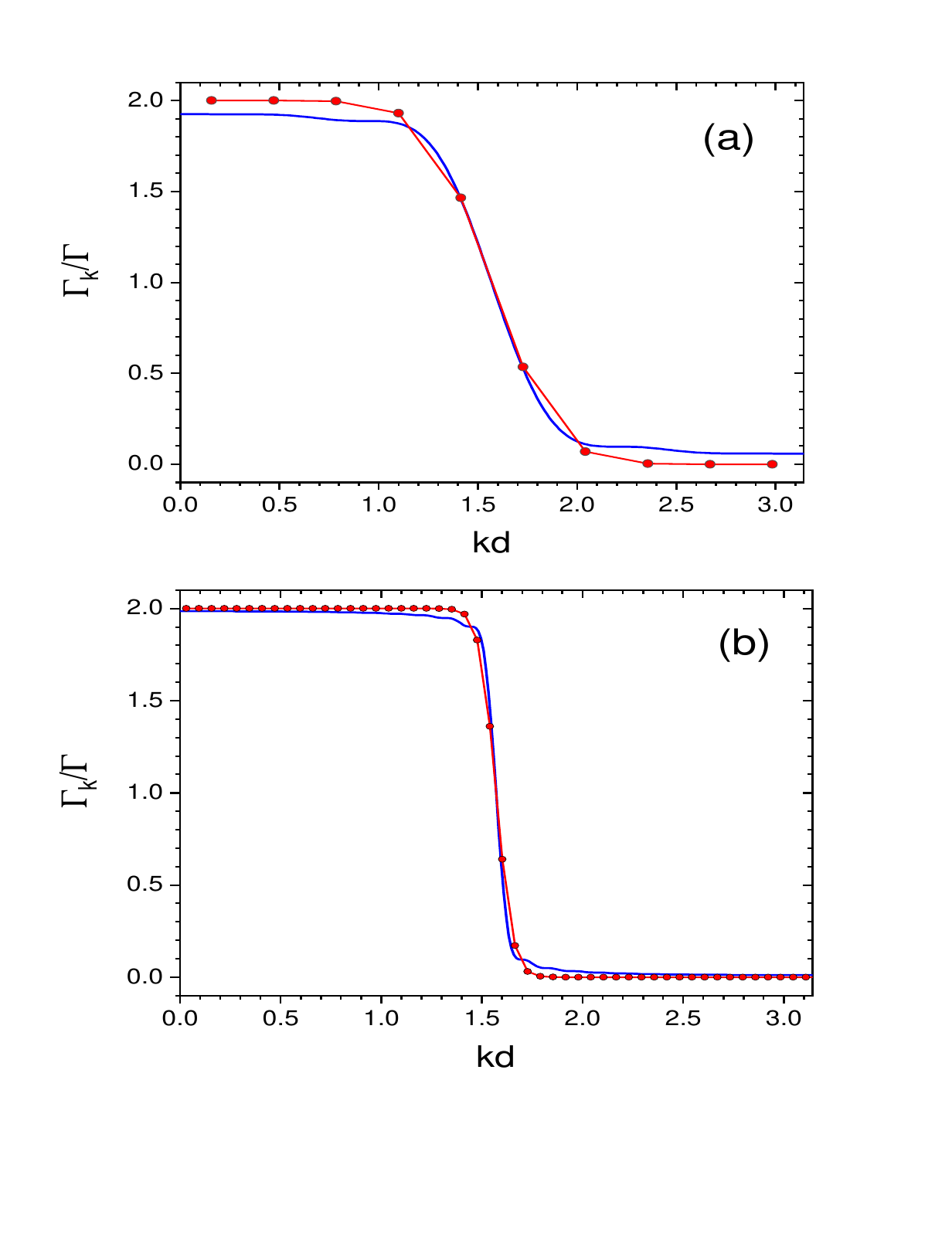}}}
      \caption{Continuous blue lines: $\Gamma_k/\Gamma$ vs $kd$ for $k_0d=\pi/2$,  obtained from Eq.(\ref{Gamma:scalar}) with (a) $N=10$ and (b) $N=100$; red circles correspond to the eigenvalues $\lambda_i$ of $\Gamma_{jm}$, ordered from the largest to the smallest, plotted as a function of $k=\pi(i-1/2)/N$.}
       \label{fig4}
    \end{figure}

Finally, it is interesting to see how $\Gamma_k$ approximates the eigenvalues spectrum of a finite chain. Fig.\ref{fig4} shows $\Gamma_k$, as calculated from Eq.(\ref{Gamma:scalar}), and the $N$ eigenvalues $\lambda_i$ of the $N\times N$ matrix $\Gamma_{jm}$, ordered from the largest to the smallest, as function of $kd=\pi(i-1/2)/N$, with $i=1,\dots,N$ for $k_0d=\pi/2$, $N=10$ (a) and $N=50$ (b).  We see that $\Gamma_k$ reproduces rather satisfactorily the features of the eigenvalues of the matrix $\Gamma_{jm}$ and describes the transition from superradiance to subradiance going from $kd=0$ to $kd=\pi$, transition which becomes sharper by increasing $N$.
       
\section{Vectorial model}

We now extend the previous analysis to the vectorial model, taking into account the polarization of the electromagnetic field. The non-Hermitian Hamiltonian is now
\begin{eqnarray}
\hat H & =&-i\frac{\hbar}{2}\sum_{\alpha,\beta}\sum_{j,j'}G_{\alpha,\beta}(\mathbf{r}_j-\mathbf{r}_{j'})\,
    \hat{\sigma}_{j,\alpha}^\dagger\hat\sigma_{j',\beta}.\label{H_vec}
\end{eqnarray}
where $\alpha,\beta=(x,y,z)$. Here  $\hat\sigma_{j,x}=(\hat\sigma_{j}^{m_J=1}+\hat\sigma_{j}^{m_J=-1})/2$, $\hat\sigma_{j,y}=(\hat\sigma_{j}^{m_J=1}-\hat\sigma_{j}^{m_J=-1})/2i$ and $\hat\sigma_{j,z}=\hat\sigma_{j}^{m_J=0}$, where 
 $\hat\sigma_{j}^{m_J}=|g_j\rangle\langle e_{j}^{m_J}|$ is the lowering operator between the ground state $|g_j\rangle$ and the three excited states $|e_{j}^{m_J}\rangle$ of the $j$th atom with quantum numbers $J=1$ and $m_J=(-1,0,1)$. The  Green function in Eq.(\ref{H_vec}) is \cite{Lehmberg1970,Needham2019}
\begin{equation}
G_{\alpha,\beta}(\mathbf{r})=\frac{3\Gamma}{2}\frac{e^{ik_0r}}{ik_0r}\left[\delta_{\alpha,\beta}-\hat n_\alpha\hat n_\beta
+\left(\delta_{\alpha,\beta}-3\hat n_\alpha\hat n_\beta\right)\left(\frac{i}{k_0 r}-\frac{1}{k_0^2 r^2}\right)
\right]
\end{equation}
with $r=|\mathbf{r}|$ and ${\hat n}_\alpha$ being the components of the unit vector $\hat{\mathbf{n}}=\mathbf{r}/r$. 
We consider the linear chain with lattice constant $d$, i.e. $\mathbf{r}_j=d(j-1)\mathbf{\hat e}_z$, with $j=1,\dots ,N$, and all the dipoles aligned with an angle $\delta$ with respect to the chain's axis, so that $\hat n_\alpha=\hat n_\beta=\cos\delta$ and
\begin{equation}
G_{\alpha,\alpha}(\mathbf{r})=\frac{3\Gamma}{2}\frac{e^{ik_0r}}{ik_0r}\left[\sin^2\delta
+(1-3\cos^2\delta)\left(\frac{i}{k_0r}-\frac{1}{k_0^2r^2}\right)\right].\label{Gaa}
\end{equation}
Notice that if the dipoles are randomly oriented, $\langle G_{\alpha,\beta}(\mathbf{r})\rangle_\Omega=0$ for $\alpha\neq\beta$ and in  Eq.(\ref{Gaa}) $\langle\cos^2\delta\rangle_\Omega=1/3$, so that  $\langle G_{\alpha,\alpha}(\mathbf{r})\rangle_\Omega=e^{ik_0r}/(ik_0r)$, i.e. we recover the scalar model. This in general is not true in disordered systems, where the short-range terms $1/r^2$ and $1/r^3$ play a role at large densities \cite{Skipetrov2013}.

\begin{figure}
       \centerline{\scalebox{0.5}{\includegraphics{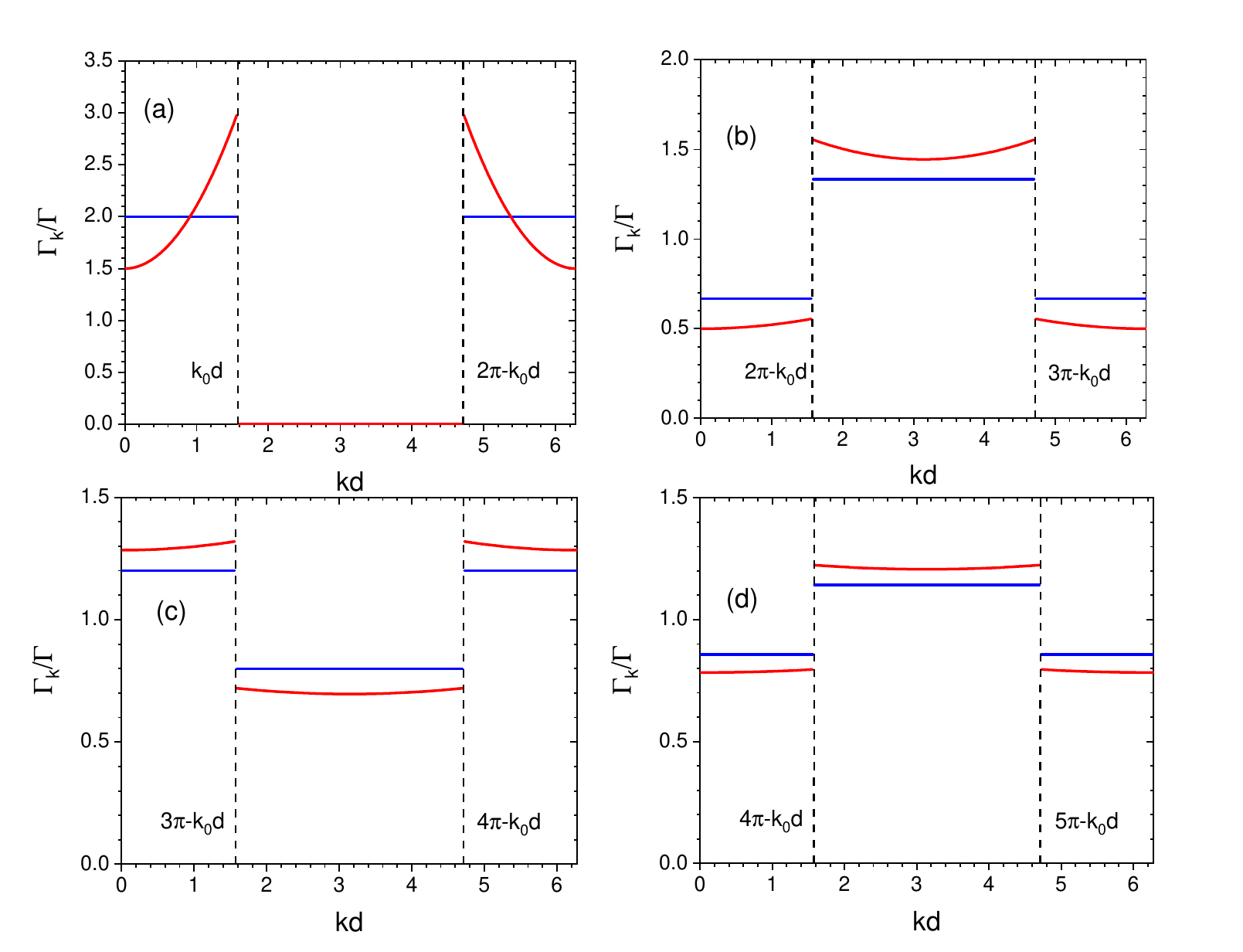}}}
      \caption{ $\Gamma_k/\Gamma$ vs $kd$ for $\delta=\pi/2$ and (a) $k_0d=\pi/2$, (b) $k_0d=3\pi/2$, (c) $k_0d=5\pi/2$ and (d) $k_0d=7\pi/2$ for an infinite chain, obtained from the vectorial model,  Eq.(\ref{gamma_vec:3}) (red line), and from the scalar model, Eq.(\ref{approx:scalar}) (blue line).}
       \label{fig5}
    \end{figure}

\begin{figure}
       \centerline{\scalebox{0.4}{\includegraphics{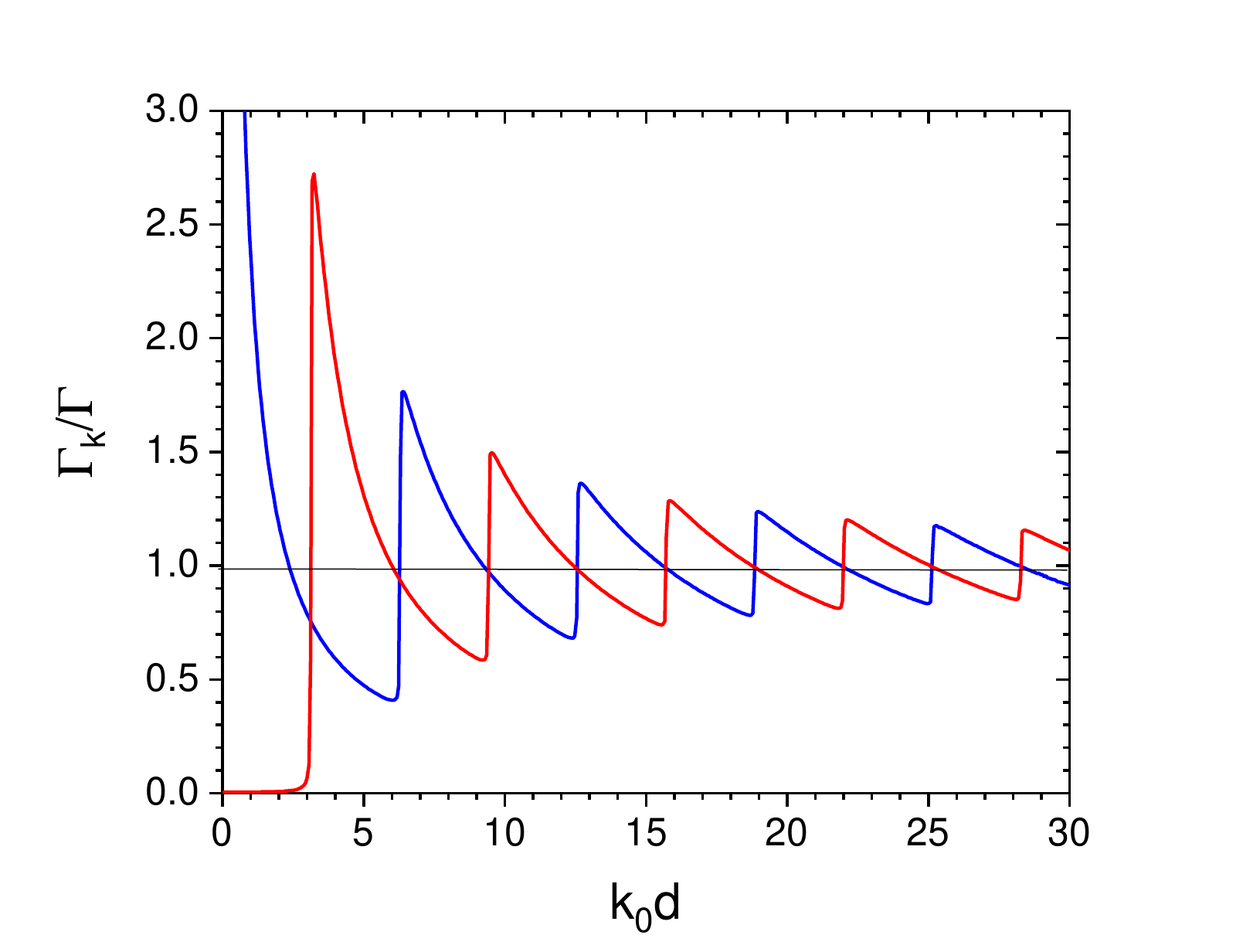}}}
      \caption{ $\Gamma_k/\Gamma$ vs $k_0d$ for $\delta=\pi/2$, for $k=0$ (blue line) and $k=\pi/d$ (red line), obtained from Eq.(\ref{gamma_vec:2}) with $N=100$.}
       \label{fig6}
    \end{figure}

The decay rate for the vectorial model is given by the real part of $G_{\alpha,\alpha}$,
\begin{eqnarray}
\Gamma^{(\delta)}(r_{jm})&=&
\frac{3\Gamma}{2}\left[\sin^2\delta j_0(k_0r_{jm})+(3\cos^2\delta-1)\frac{j_1(k_0r_{jm})}{k_0r_{jm}}\right]\label{gamma_vec:general}
\end{eqnarray}
where $j_0(x)=\sin x/x$ and $j_1(x)=\sin x/x^2-\cos x/x$ are the spherical Bessel functions of order $n=0$ and $n=1$. As before, it is possible to write $\Gamma^{(\delta)}(r_{jm})$ as angular average of the radiation field emitted between the two atoms. By using the identities
\begin{eqnarray}
\langle e^{ix\cos\theta}\rangle_\Omega &=&\frac{1}{4\pi}\int_0^{2\pi}d\phi\int_0^\pi \sin\theta e^{ix\cos\theta}d\theta=j_0(x)\\
\langle \cos^2\theta e^{ix\cos\theta}\rangle_\Omega &=&\frac{1}{4\pi}\int_0^{2\pi}d\phi\int_0^\pi \sin\theta\cos^2\theta e^{ix\cos\theta}d\theta=
j_0(x)-2\frac{j_1(x)}{x}.
\end{eqnarray}
we can write
\begin{eqnarray}
\Gamma^{(\delta)}(r_{jm})&=&\frac{3\Gamma}{4}\left\{\frac{1+\cos^2\delta}{2}\langle E_j E_m^*\rangle_\Omega +\frac{1-3\cos^2\delta}{2}
\langle \cos^2\theta E_jE_m^*\rangle_\Omega+\mathrm{c.c.}
\right\}
\end{eqnarray}
where $E_j$ has been defined in the previous section. The collective decay rate of the dipoles oriented with the angle $\delta$ is
\begin{equation}
\Gamma_k^{(\delta)}=\frac{1}{N}\sum_{j=1}^N\sum_{m=1}^N\Gamma^{(\delta)}(r_{jm})e^{ikd(j-m)}
\end{equation}
with $k\in[0,2\pi/d]$. Then,
\begin{eqnarray}
  \Gamma_k^{(\delta)} &=&\frac{3\Gamma}{4N}\left[(1+\cos^2\delta)\langle|F_k(\theta)|^2\rangle_\Omega+
  (1-3\cos^2\delta)\langle\cos^2\theta|F_k(\theta)|^2\rangle_\Omega\right]
\end{eqnarray}
where $|F_k(\theta)|^2$ is defined in Eq.(\ref{Fk}). By evaluating the angular average and using the approximation (\ref{approx}),
\begin{eqnarray}
  \Gamma_k^{(\delta)} &=&\frac{3\Gamma N}{2k_0d}\sum_{m=-\infty}^{+\infty} \int_{(k-k_0)d/2}^{(k+k_0)d/2}
  \left[\sin^2\delta+\frac{1}{2}(1-3\cos^2\delta)\frac{(kd-2t)^2-(k_0d)^2}{(k_0d)^2}
  \right]\nonumber\\
  &\times&\mathrm{sinc}^2\left[\left(t-m\pi\right)N\right]dt.\label{gamma_vec:2}
\end{eqnarray}
In the limit $N\rightarrow\infty$, using (\ref{Dirac}),
\begin{eqnarray}
  \Gamma_k^{(\delta)} &=&\frac{3\Gamma\pi}{2k_0d}\sum_{m=-\infty}^{+\infty}
  \left\{\sin^2\delta+\frac{1}{2}(1-3\cos^2\delta)\frac{(kd-2\pi m)^2-(k_0d)^2}{(k_0d)^2}
  \right\}\nonumber\\
  &\times&\Pi[2m\pi-k_0d<kd<2m\pi+k_0d],\label{gamma_vec:3}
\end{eqnarray}
which is the Fourier transform of the decay rates for an infinite chain.
We observe that, for $k_0d<\pi$, $\Gamma_k^{(\delta)}$ is still zero in the interval $k_0<k<2\pi/d-k_0$ (full subradiance). Instead, 
$\Gamma_k^{(\delta)}$  it no more uniform in $k$ as in the scalar model. In particular, for
$k_0d<\pi$ and in the intervals $0<k<k_0$ and $2\pi/d-k_0<k<2\pi/d$,
\begin{eqnarray}
  \Gamma_k^{(\delta)} &=&\frac{3\Gamma\pi}{2k_0d}
  \left\{\sin^2\delta+\frac{1}{2}(1-3\cos^2\delta)\left(\frac{k^2}{k_0^2}-1\right)
  \right\}\label{Gamma:delta}
\end{eqnarray}
Eq.(\ref{Gamma:delta}) is in agreement with the results of ref.\cite{Asenjo2017}.
In Fig.\ref{fig5} we compare the results of the vectorial and the scalar models for an infinite chain, for the same values of $k_0d$ as in Fig.\ref{fig1} and for $\delta=\pi/2$. We observe that increasing $k_0d$ the differences between the scalar and the vectorial models become less important and the spectrum becomes more flat. Also the behavior of 
$\Gamma_k^{(\delta)}$ vs $k_0d$ for $k=0$ and $k=\pi/d$, shown in Fig.\ref{fig6}, is similar to that obtained from the scalar model (see Fig.\ref{fig2}). We observe that $\Gamma_k^{(\delta)}$ reduces to the scalar model's value, $\Gamma_k$, for the special angle $\delta=\arccos(1/\sqrt{3})$.

\begin{figure}
      \centerline{\scalebox{0.4}{\includegraphics{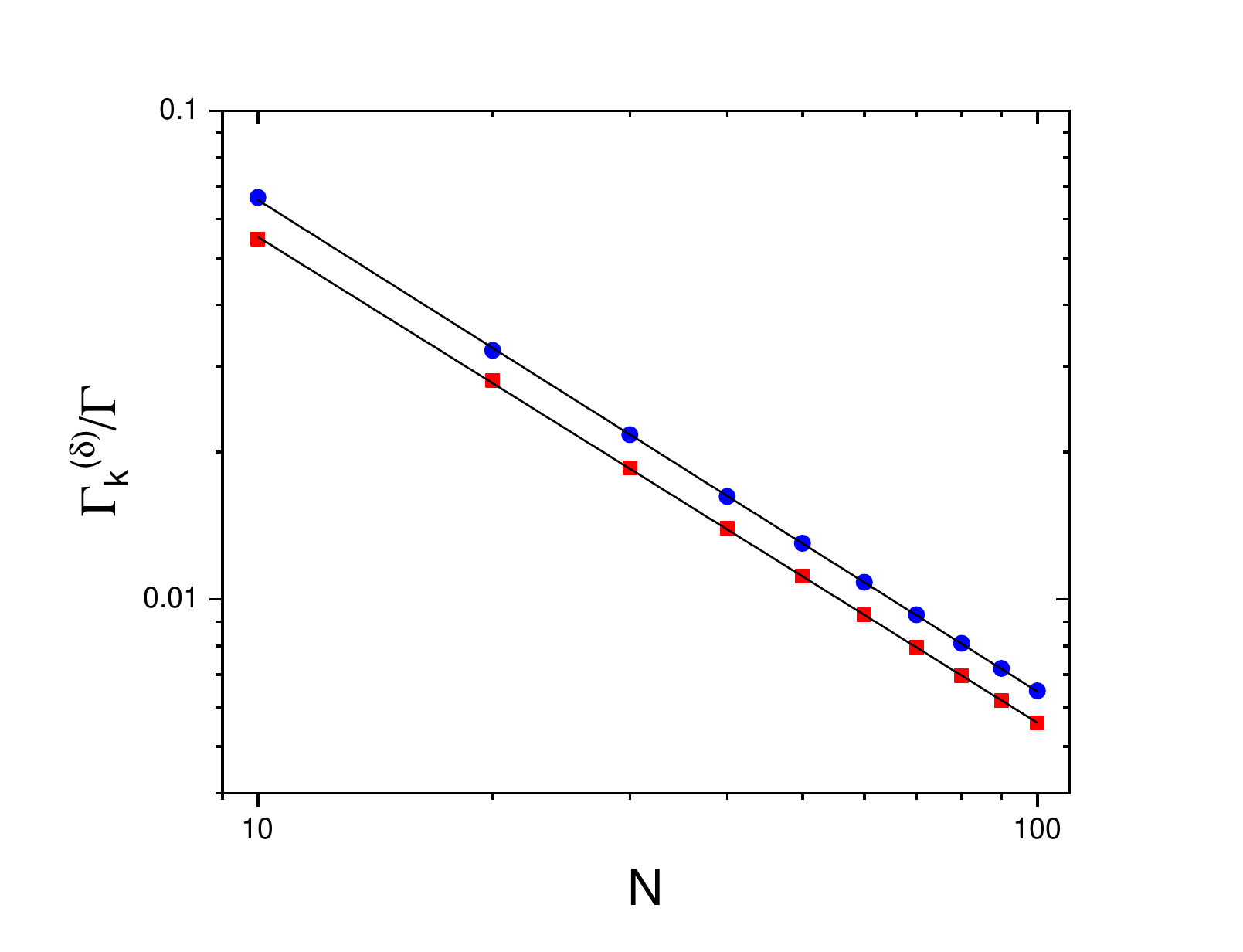}}}
      \caption{ Scaling with atom number $N$ of the subradiant decay rate $\Gamma_k^{(\delta)}/\Gamma$ for $k=\pi/d$ and $k_0d=\pi/2$. A fit for large $N$ yields $\Gamma_k^{(\delta)}\sim 1/N$, both for perpendicular ($\delta=\pi/2$, circles) and parallel, ($\delta=0$, squares), polarization.}
       \label{fig7}
\end{figure}

Finally, as done for the scalar model, we can obtain an approximated expression of $\Gamma_k^{(\delta)}$ for a finite lattice, substituting the function $\mathrm{sinc}^2(x)$ with the Lorentzian function $1/(1+x^2)$ in Eq.(\ref{gamma_vec:2}) and solving the integral. A long but straightforward calculation yields:
\begin{eqnarray}
  \Gamma_k^{(\delta)}(x,a)&=&\frac{3\Gamma}{2a}\sum_{m=-\infty}^{+\infty}\left\{
  \left[\sin^2\delta+\frac{1}{2a^2}(1-3\cos^2\delta)[(x-2\pi m)^2-a^2]\right]\right.\nonumber\\
  &\times &\left.(\arctan b_m-\arctan a_m)\right.\nonumber\\
  &+&\frac{2}{Na}(1-3\cos^2\delta)-\frac{1}{Na^2}(1-3\cos^2\delta)(x-2\pi m)\ln\frac{1+b_m^2}{1+a_m^2}\nonumber\\
  &-&\left. \frac{2}{N^2a^2}(1-3\cos^2\delta)(\arctan b_m-\arctan a_m)\right\}
  \label{gamma_vec:arctan}
\end{eqnarray}
where we defined $x=kd$, $a=k_0d$, $a_m=(x-2\pi m-a)N/2$ and $b_m=(x-2\pi m+a)N/2$. A numerical analysis shows that the approximated expression (\ref{gamma_vec:arctan}) reproduces satisfactorily the exact solution (\ref{gamma_vec:2}).
We have verified the $1/N$ scaling of subradiance by evaluating numerically the exact solution (\ref{gamma_vec:2}) for $kd=\pi$ and $k_0d=\pi/2$ and two different polarization directions, $\delta=\pi/2$ and $\delta=0$, shown in Fig.\ref{fig7}. The linear fit confirms the  subradiant law  $\Gamma_{\mathrm{sub}}\sim 1/N$, independently from the polarization direction.

\section{Conclusions}

We have presented a different approach to the study of the cooperative decay in a one-dimensional chain of $N$ atoms in the single-excitation configuration. We have defined a collective function  $\Gamma_k$, containing the phase information among the emitters. In particular, it reduces to the Fourier transform of the decay rates in the case of an infinite chain.
For a finite chain, it gives an intuitive interpretation of superradiance and subradiance as positive or negative interference of the emission by $N$ atoms in the chain. More specifically, $\Gamma_k$ is proportional to the imaginary part of the expectation value of the non-Hermitian Hamiltonian operator, evaluated on a generalized Dicke state of $N$ atoms with nearest-neighbor phase difference $kd$, where $d$ is the lattice constant: superradiance and subradiance  occurs for $k=0$ and $k=\pi/d$,  respectively, however only  for $d/\lambda_0<1/2$. 
Our analytic results are in agreement with the numerical  results of \cite{Bettles2016,Masson2020}. More importantly, this approach can be complementary to the study of the discrete eigenvalue problem, generally numerically limited by the size of $N$. We have derived an explicit expression of $\Gamma_k$, both for the scalar and vectorial light models. From it, subradiance shows a dependence as $1/N$, similar to that predicted and observed in disordered systems \cite{Bienaime2012,Guerin2016}. This result seems to be in odd with the analysis of ref.\cite{Asenjo2017}, where a dependence as $1/N^3$ is found for the most subradiant eigenvalues. However, that scaling refers to discrete eigenvalues, whose behavior depends in general on the microscopic details, such for instance the polarization of the atoms. On the contrary,  $\Gamma_k$ is a mean-field quantity, obtained as an expectation value of the non-Hermitian Hamiltonian operator, evaluated over a collective state with a well-defined phase relationship between nearest-neighbor atoms. In this sense, $\Gamma_k$ could be associated to the measurement of the average subradiant decay in a finite linear chain, rather than the less accessible  eigenvalues of the non-Hermitian Hamiltonian.
It will be interesting to extend this approach of the study of the cooperative decay from one-dimensional chains to two- and three- dimensional finite arrays, eventually also in the presence of disorder.

\acknowledgments{The author thank B. Olmos and S. Olivares for helpful discussions.}

\end{document}